\documentclass[aps,prl,reprint,superscriptaddress]{revtex4-1}
\pdfoutput=1
 
\usepackage{graphicx}
\usepackage{float}
\usepackage{floatflt}
\usepackage{lipsum}
\begin{document}


\title{Symmetry-controlled second-harmonic generation in a plasmonic waveguide}

\author{Tzu-Yu Chen}
\affiliation
{Institute of Photonics Technologies, National Tsing Hua University, Hsinchu 30013, Taiwan}
\affiliation{International Intercollegiate PhD Program, National Tsing Hua University, Hsinchu 30013, Taiwan}

\author{Julian Obermeier}
\affiliation{Department of Physics, University of Bayreuth, 95440 Bayreuth, Germany}

\author{Thorsten Schumacher}
\affiliation{Department of Physics, University of Bayreuth, 95440 Bayreuth, Germany}

\author{Fan-Cheng Lin}
\affiliation{Department of Chemistry, National Tsing Hua University, Hsinchu 30013, Taiwan}  

\author{Jer-Shing Huang}
\affiliation{Leibniz Institute of Photonic Technology, 07745 Jena, Germany}
\affiliation{Department of Electrophysics, National Chiao Tung University, Hsinchu 30010, Taiwan}
\affiliation{Research Center for Applied Sciences, Academia Sinica, Taipei 115-29, Taiwan}

\author{Markus Lippitz}
\email{markus.lippitz@uni-bayreuth.de}
\affiliation{Department of Physics, University of Bayreuth, 95440 Bayreuth, Germany}

\author{Chen-Bin Huang}
\email{robin@ee.nthu.edu.tw}
\affiliation
{Institute of Photonics Technologies, National Tsing Hua University, Hsinchu 30013, Taiwan}
\affiliation{International Intercollegiate PhD Program, National Tsing Hua University, Hsinchu 30013, Taiwan}
\affiliation{Research Center for Applied Sciences, Academia Sinica, Taipei 115-29, Taiwan}



\begin{abstract}
	A new concept for second-harmonic generation (SHG) in an optical nanocircuit is proposed. We demonstrate both theoretically and experimentally that the symmetry of an optical mode alone is sufficient to allow SHG even in centro-symmetric structures made of centro-symmetric material. The concept is realized using a plasmonic two-wire transmission-line (TWTL), which simultaneously supports a symmetric and an anti-symmetric mode. We first confirm the generated second-harmonics belong only to the symmetric mode of the TWTL when fundamental excited modes are either purely symmetric or anti-symmetric. We further switch the emission into the anti-symmetric mode when a controlled mixture of the fundamental modes is excited simultaneously. 
Our results open up a new degree of freedom into the designs of nonlinear optical components, and should pave a new avenue towards multi-functional nanophotonic circuitry. 
\end{abstract}


\maketitle


\begin{figure}
\includegraphics[width=\columnwidth]{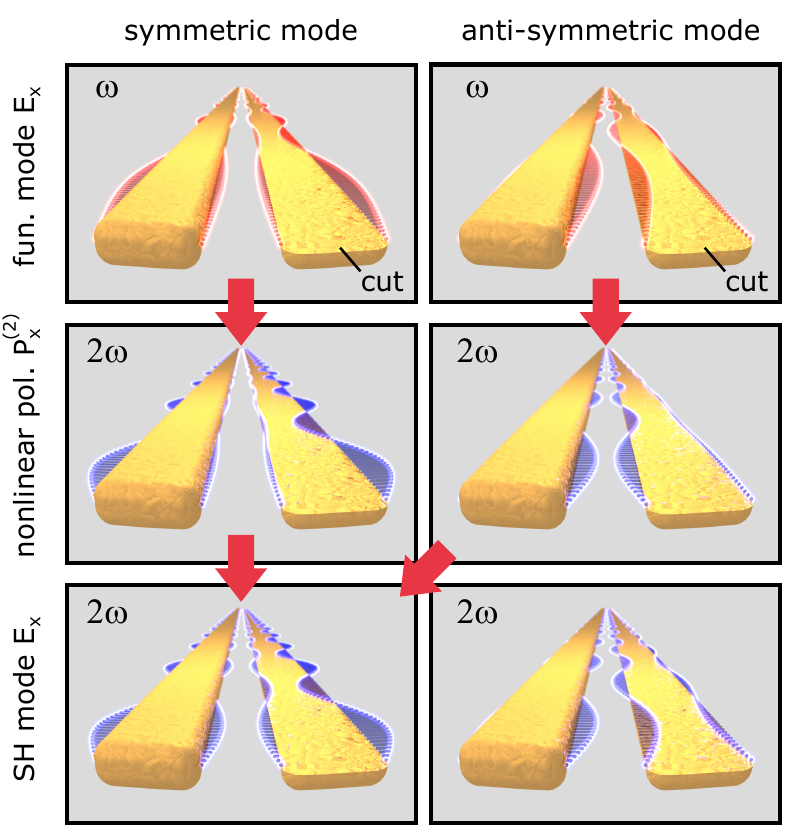}
\caption{\textbf{Schematics of second-harmonic generation within a fully symmetric waveguide.} A two-wire transmission line supports two optical modes of different symmetry as depicted by the electric field vectors along the middle of the waveguide (top). The nonlinear polarization responsible for second-harmonic generation is proportional to the square of the field along the surface normal (mid). This polarization launches an optical wave in case a suitable second-harmonic (SH) mode is present. In the case of the waveguide, the symmetric mode is symmetry-allowed for both fundamental modes (bottom). In the figures, the top part of the right transmission line is removed to facilitate revealing the field components pointing into the  nanowire.
 \label{fig:idea}}
\end{figure}


As documented in all nonlinear optics textbooks, second-harmonic generation (SHG) from a bulk material requires a non-centro-symmetric crystal structure. Typical noble metals exploited for plasmonics, such as gold, silver and aluminum, are centro-symmetric and do not allow SHG in their bulk form. A strategy to circumvent this limitation is to make use of the non-vanishing second-order susceptibility at the surfaces, where the symmetry is automatically broken \cite{Brown1965,Bloembergen1968,Simon1974,Makitalo2011}. However, when measured in the far-field, despite local SHG is permitted at the interface, contributions of two opposing surface elements cancel out provided the nanostructure is exhibiting centro-symmetry. 
%
%
Past efforts in plasmonic-assisted SHG  circumvented this cancellation through designing asymmetric nanostructures where  the amplitude of one surface contribution overwhelms that of the opposing surface. 
It is interesting to note that prior plasmonic-assisted SHGs were dominated by localized surface plasmons, i.e., the metallic nanostructures act as optical antennas \cite{Zhang2011,Konishi2014,OBrien2015,Czaplicki2015,Celebrano2015,Gennaro2016,Gomez-Tornero2017,Yang2017,Chervinskii2018}. On contrary, rare research attention has been paid to nonlinear frequency conversions using propagating plasmons \cite{Chen1979,Viarbitskaya2015,deHoogh2016,Li2017,Lan:2015bi}. Despite an early demonstration of SHG in a two-dimensional metallic thin film \cite{Chen1979}, to our best knowledge, only three works have employed three-dimensionally confined surface plasmon polaritons (SPPs) \cite{Viarbitskaya2015,deHoogh2016,Li2017}. We note here in Refs.~\cite{Viarbitskaya2015} and \cite{Li2017}, the generated second-harmonic signals were no longer guided by the plasmonic structures, but radiating into the far-field.

A key ingredient of our approach is a propagating plasmonic waveguide mode. In this letter, we demonstrate emission of second harmonic light which is thought to be forbidden, as the plasmonic nanocircuit is fully centro-symmetric in geometry and material.
The structure we use is a two-wire transmission line (TWTL) and consists of two parallel gold nanowires of identical size and shape~\cite{Geisler2013,Wu2017,Dai2014}. Although the waveguide is still centro-symmetric, it supports a symmetric and an anti-symmetric mode (Fig.~\ref{fig:idea} top), similar to the bonding and anti-bonding orbital in a hydrogen molecule or the two eigenmodes of a coupled pendulum. However, due to the absence of any resonance condition, both modes are broadband and cover large parts of the optical and near-infrared spectrum. We label the two modes  by the symmetry of the normal components of the electric field, which agrees with the symmetry of the charge distribution. During the propagation of the fundamental field, a nonlinear polarization $\mathbf{P}^{(2)}(2 \omega)$ is generated locally (Fig.~\ref{fig:idea} mid). It can emit into the waveguide modes available at the second-harmonic frequency $2 \omega$ (Fig.~\ref{fig:idea} bottom). 
While the fundamental ($\lambda = 1560$~nm) and second harmonic fields ($\lambda = 780$~nm) are computed with a finite element solver (Comsol), the nonlinear polarization is calculated as \cite{Makitalo2011}
\begin{equation}
P^{(2)}_n = \epsilon_0 \,  \chi^{(2)}_{nnn} \, E_n^2   \quad ,
\label{eq:p2_nnn}
\end{equation}
where $n$ denotes the vector component normal to the surface. In second-harmonic generation at gold surfaces the tensor component $\chi^{(2)}_{nnn}$ dominates, i.e., only the vector components along the surface normal $n$ enter \cite{Makitalo2011}.

The coupling efficiency $\eta$ of the nonlinear polarization $\mathbf{P}^{(2)}$ to the waveguide modes, at the second harmonic, can be calculated as \cite{OBrien2015,Roke:2004cf}
\begin{equation}
 \eta = \left| \int_{\partial A} \mathbf{P}^{(2)}(2 \omega)\, \cdot \,  \mathbf{E}(2 \omega)^* \, ds \right |^2
\label{eq:eta}
\end{equation}
where the integral runs over the  surface of the waveguide. The integral reduces to a line integral along the circumference in case phase matching and damping is neglected, as possible for waveguides of a fixed length. In case of symmetric and anti-symmetric  field distributions the nonlinear polarization  is always  fully symmetric. In consequence, only emission into the symmetric mode is allowed, while the integral for the anti-symmetric mode vanishes. In analogy, this is what forbids second-harmonic emission from a small plasmonic sphere, as free space modes are anti-symmetric upon point inversion. Numerical calculations taking all details into account support the symmetry argument (see supplementary material).

\begin{figure}
\includegraphics[width=\columnwidth]{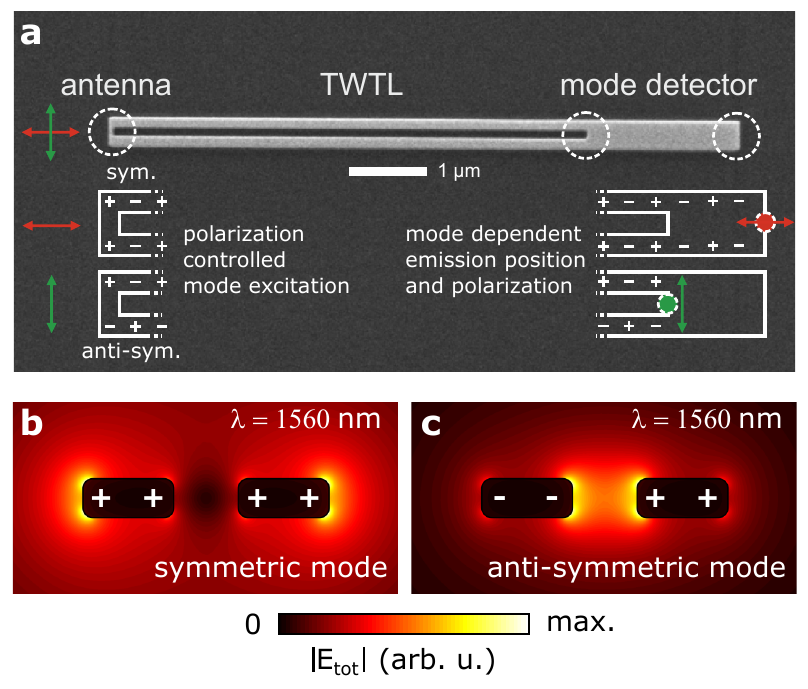}
\caption{\textbf{Experimental realization of a plasmonic two-wire transmission line supporting two waveguide modes}. (a) SEM micrograph of the structure, fabricated by focused ion-beam milling of a crystalline gold flake. A thin connection between the wires serves as incoupling antenna. A region without gap between the wires acts as mode detector. Depending on the polarization direction  of the laser focused in the antenna part, either the symmetric (b) or the anti-symmetric mode (c) is launched. The latter is localized between the two wires and scattered out on the left end of the mode detector. The symmetric mode travels until the right end of the mode detector. The emission has the same polarization direction as the excitation beam.
\label{fig:coupler}}
\end{figure}

We now set out to experimentally demonstrate these predictions. The waveguide consists of two identical single-crystalline gold wires of about 100 nm width and distance (Fig.~\ref{fig:coupler}a).
A plasmonic nanoantenna that is connected to the transmisison line (left side of structure) allows to independently excite the two eigenmodes of the waveguide by controlling the polarization direction of the incoming laser beam. Here, the polarization along the antenna (vertical axis) leads to a dipolar charge distribution and launches the anti-symmetric mode (inset Fig.~\ref{fig:coupler}a). In case of the horizontal polarization, the antenna part acts as scattering edge launching the symmetric mode. For detection, we make use of the different modal distribution of the electric field-amplitude. The anti-symmetric mode is confined between the wires while the symmetric mode has a higher field amplitude outside the wire pair (Fig.~\ref{fig:coupler}b,c). This allows us to use a two-wire/single-wire interface (right side of structure) as mode detector \cite{Geisler2013,Dai2014,Rewitz2014}. The emitted light has the identical polarization that was necessary to excite the mode.

\begin{figure}
\includegraphics[width=\columnwidth]{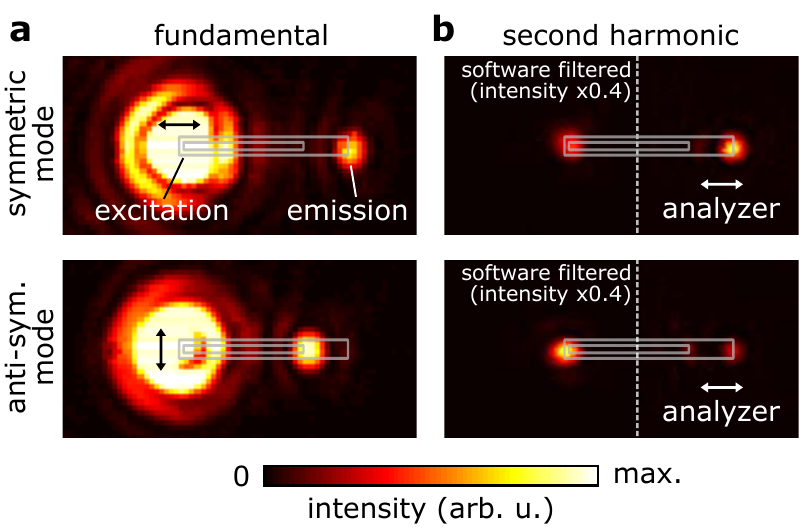}
\caption{\textbf{Emission from the mode detector signals generation of second-harmonic in the  symmetric mode, independent of the launched fundamental mode.} (a)  Imaging the sample at the fundamental wavelength shows next to the strong reflection of the incoming laser beam a second spot either at the left or at the right end of the mode detector. Depending on the input polarization (black arrows) either the symmetric or anti-symmetric mode is launched. (b) Imaging at the second harmonic shows also emission from the antenna itself, but independent of the launched fundamental mode always emission from the symmetric waveguide mode at the right end of the mode detector. The weak spot at the left end of the mode detector is a background signal from locally generated second harmonic (see supplementary material). For better visibility, the displayed intensity of excitation sides  is reduced by 60\%. 
  \label{fig:images}}
\end{figure}

The experimental results for linear and nonlinear propagation are shown in Figs.~\ref{fig:images}a and b. When exciting the antenna with horizontal polarization, the symmetric mode is launched leading to fundamental emission at the right end of the mode detector (top left panel). Rotating the polarization by $\pi / 2 $, the excited mode changes and consequently also the emission point (bottom left panel). 
In the second harmonic images Fig.~\ref{fig:images}b the analyzer selects the emission from the symmetric mode at the right end of the mode detector. A discussion of the other polarization direction is presented in the supplementary material. 
We find locally generated second-harmonic emission from the input antenna, independent on the fundamental excitation polarization. In contrast to other publications, the waveguide itself appears dark, demonstrating high surface quality~\cite{deHoogh2016}.
Most important, we find second-harmonic emission for both excitation polarization directions at the right end of the mode detector, i.e., stemming from the symmetric mode only. The excitation of the anti-symmetric fundamental mode (bottom right panel) leads to a slightly lower emission intensity, in agreement with our numerical simulations, due to a reduced mode overlap.
Regardless of which pure fundamental mode is excited, we always generate second-harmonic during propagation in the waveguide that is emitted into the symmetric mode at frequency  $2\omega$. This process is symmetry-allowed and takes place although both the material as well as the structure is centro-symmetric. \newline

\begin{figure}[h!]
\includegraphics[width=\columnwidth]{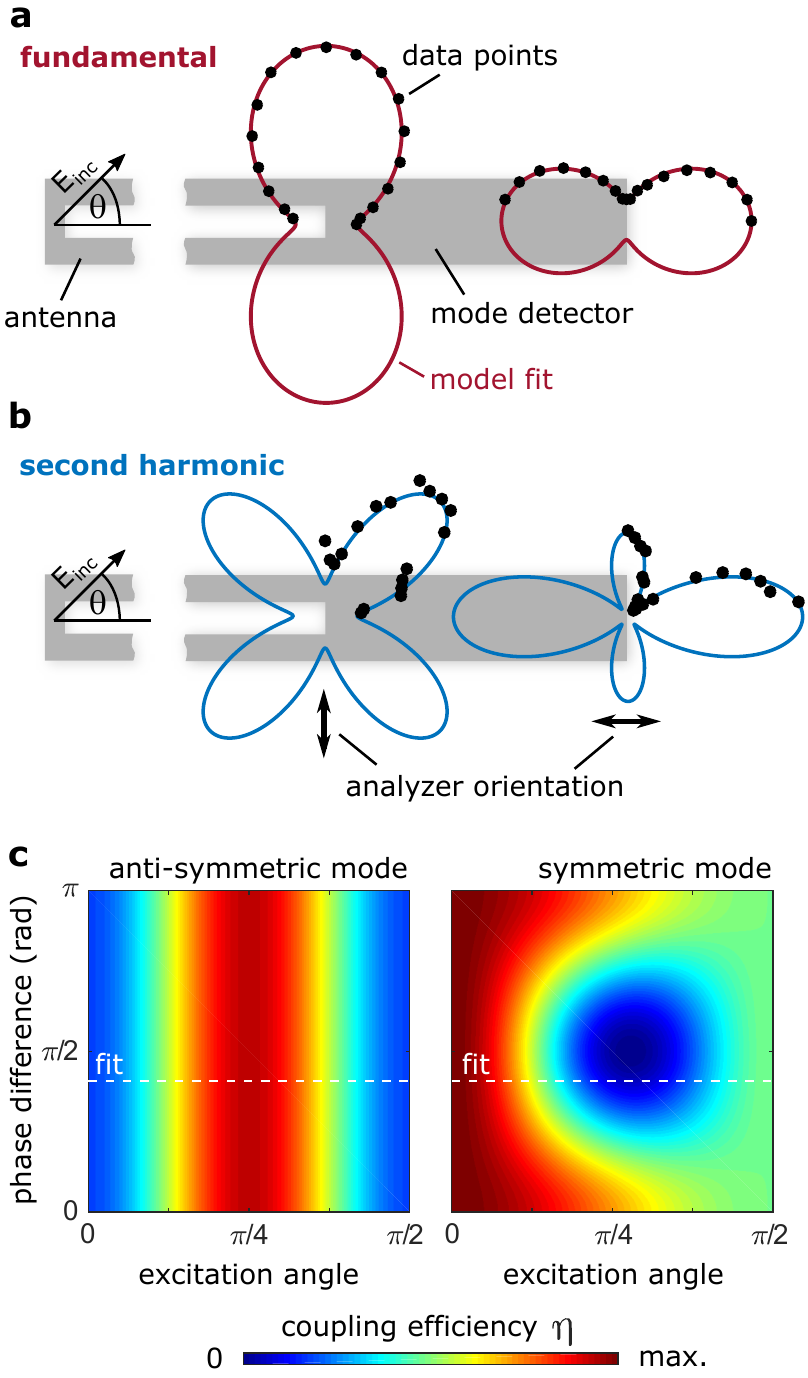}
\caption{\textbf{Coupling efficiency of the second-order nonlinear polarization to the symmetric and anti-symmetric mode of the waveguide.} (a) Fundamental emission intensity at the two mode-detector spots, as a function of the excitation beam polarization angle $\theta$. The linear response follows the expected behavior. (b) Observing the nonlinear emission at the mode detector in the expected emission polarization direction shows a more complex behavior. It can be modeled taking the coherent superposition of symmetric and anti-symmetric fundamental mode into account. From the fit parameters ($r = 0.83$, $\phi = 50.1 \deg$) we can predict the full excitation polarization and phase dependence of the second harmonic emission intensity of the symmetric and anti-symmetric mode (c). 
\label{fig:polar}}
\end{figure}

Obviously the fundamental harmonic excitation and subsequently the second harmonic emission into purely symmetric and anti-symmetric modes are special cases of a  more general behavior, that we want to study in the following. In general, each superposition of fundamental modes may be launched, leading to a nonlinear mode mixing during propagation and emission from different second harmonic modes.
In case of our waveguide, the fundamental field can be written as a coherent superposition of the symmetric ($\mathbf{E}_s$) and anti-symmetric ($\mathbf{E}_{as}$) waveguide mode:
\begin{equation}
\mathbf{E}_{tot} = c_x \,   \mathbf{E}_s \, + \, c_{y} \, r  \exp(i \phi) \, \mathbf{E}_{as} \quad ,
\label{eq:Etot}
\end{equation}

with $c_x = \cos \theta$, $c_{y} = \sin \theta$  the $x$ and $y$-polarized excitation field components of the fundamental laser beam. The parameters $r$ and $\phi$ take into account that the antenna coupling efficiency differs in amplitude and phase between the two modes. When observing  the fundamental wavelength, the mode detector projects the total field $\mathbf{E}_{tot} $ again on the two eigenmodes. We observe the expected $\cos^2 \theta$ ($\sin^2 \theta$) dependence for the symmetric (anti-symmetric) mode intensity, as depicted by the polar representation in Fig.~\ref{fig:polar}a. The behavior gets more complex when observing the second harmonic emission (Fig.~\ref{fig:polar}b). For the symmetric mode we find a deep minimum near 45 degree between the differing peak values at $0$ and $90$~degree that were already discussed in Fig.~\ref{fig:images}b. Moreover, the anti-symmetric mode shows emission with a peak at 45 degree linear input polarization of the fundamental mode.

These results can be fully recovered by the model that we already introduced, taking only the dominating normal components of the fields into account.  The total fundamental field ${E}_{tot} $ gives rise to a nonlinear polarization $P^{(2)}_n$ (eq.~\ref{eq:p2_nnn}) that is coupled with certain efficiency $\eta$ (eq.~\ref{eq:eta}) into the target mode at second harmonic frequency. The nonlinear polarization can  be separated into a symmetric and anti-symmetric part,
\begin{equation}
{P}_{n}^{(2)} \propto 
\underbrace{ c_x^2 \, {E}_s^2 +  c_y^2  \, r^2 \exp(2 i  \phi)  \, {E}_{as}^2}_{\text{symmetric}} + 
 \underbrace{  c_x \,  c_y \, r \exp(i  \phi) \,  {E}_s {E}_{as}}_{\text{anti-symmetric}}
\label{eq:model}
\end{equation}
As both ${E}_s^2$ and ${E}_{as}^2$ describe symmetric field distributions, only the cross term contributes to second-harmonic generation in the anti-symmetric mode. The model is fitted to the full dataset using  $r$ and $\phi$ as free parameters as well as  scaling parameters for the absolute intensity. The unknown exact spatial shape of the modes ${E}_s$ and ${E}_{as}$ for the coupling efficiency integral  (eq.~\ref{eq:eta}), as well as the influence of the second harmonic phase matching can be absorbed in the free parameters $r$ and $\phi$ (see supplementary material). 

In Fig.~\ref{fig:polar}c we explore the dependence of the mode intensities on the relative phase $\phi$ and the polarization direction $\theta$. The anti-symmetric mode is excited at the second harmonic if both modes are present in the fundamental field. It scales with the product of both mode amplitudes,  but it does not depend on the relative phase $\phi$. The symmetric target mode in contrast shows interference between the nonlinear polarization that is generated by each fundamental mode. For a relative phase $\phi$ around $\pi/2$ this interference is  destructive, causing the minimum around $\theta = 50.1$ degree. The position and depth of this minimum depends on the value of $r$. This makes it possible to switch the second harmonic emission fully from the symmetric to the anti-symmetric mode and back again by turning the fundamental polarization direction.

To conclude, the symmetry of plasmonic waveguide modes offers new opportunities not available in far-field optics. This allows us to demonstrate a novel scheme for nanoscale second-harmonic generation. Although both material and structure are fully symmetric, the propagating plasmon in a  smooth two-wire transmission-line   generates second-harmonic in the waveguide modes. When the traveling fundamental  field is either purely symmetric or anti-symmetric, only the symmetric second-harmonic mode is excited. A  coherent superposition of the fundamental modes allows to control the amplitude ratio in the emitting modes, up to fully switching to the anti-symmetric second-harmonic mode. Our finding could be extended to any generalized waveguide systems where modes of suitable symmetry are supported. Our results opens up a completely new degree of freedom into the designs of nonlinear optical components, and should path a new avenue towards multi-functional nanophotonic circuitry.

The authors thank the support from the Ministry of Science and Technology in Taiwan under Grants MoST 103-2112-M-007-017-MY3 and MoST 106-2112-M-007-004-MY3.

F.C.L. and J.S.H. prepared the samples.
C.B.H. and T.Y.C.  designed  and performed the experiments. 
J.O., T.S. and M.L. performed numerical simulations and data processing.
C.B.H., J.O., T.S., and M.L. wrote the manuscript. All authors commented on the manuscript.
T.Y.C. and J.O. contributed equally.

%


\begin{thebibliography}{23}%
	\makeatletter
	\providecommand \@ifxundefined [1]{%
		\@ifx{#1\undefined}
	}%
	\providecommand \@ifnum [1]{%
		\ifnum #1\expandafter \@firstoftwo
		\else \expandafter \@secondoftwo
		\fi
	}%
	\providecommand \@ifx [1]{%
		\ifx #1\expandafter \@firstoftwo
		\else \expandafter \@secondoftwo
		\fi
	}%
	\providecommand \natexlab [1]{#1}%
	\providecommand \enquote  [1]{``#1''}%
	\providecommand \bibnamefont  [1]{#1}%
	\providecommand \bibfnamefont [1]{#1}%
	\providecommand \citenamefont [1]{#1}%
	\providecommand \href@noop [0]{\@secondoftwo}%
	\providecommand \href [0]{\begingroup \@sanitize@url \@href}%
	\providecommand \@href[1]{\@@startlink{#1}\@@href}%
	\providecommand \@@href[1]{\endgroup#1\@@endlink}%
	\providecommand \@sanitize@url [0]{\catcode `\\12\catcode `\$12\catcode
		`\&12\catcode `\#12\catcode `\^12\catcode `\_12\catcode `\%12\relax}%
	\providecommand \@@startlink[1]{}%
	\providecommand \@@endlink[0]{}%
	\providecommand \url  [0]{\begingroup\@sanitize@url \@url }%
	\providecommand \@url [1]{\endgroup\@href {#1}{\urlprefix }}%
	\providecommand \urlprefix  [0]{URL }%
	\providecommand \Eprint [0]{\href }%
	\providecommand \doibase [0]{http://dx.doi.org/}%
	\providecommand \selectlanguage [0]{\@gobble}%
	\providecommand \bibinfo  [0]{\@secondoftwo}%
	\providecommand \bibfield  [0]{\@secondoftwo}%
	\providecommand \translation [1]{[#1]}%
	\providecommand \BibitemOpen [0]{}%
	\providecommand \bibitemStop [0]{}%
	\providecommand \bibitemNoStop [0]{.\EOS\space}%
	\providecommand \EOS [0]{\spacefactor3000\relax}%
	\providecommand \BibitemShut  [1]{\csname bibitem#1\endcsname}%
	\let\auto@bib@innerbib\@empty
	\bibitem [{\citenamefont {Brown}\ \emph {et~al.}(1965)\citenamefont {Brown},
		\citenamefont {Parks},\ and\ \citenamefont {Sleeper}}]{Brown1965}%
	\BibitemOpen
	\bibfield  {author} {\bibinfo {author} {\bibfnamefont {F.}~\bibnamefont
			{Brown}}, \bibinfo {author} {\bibfnamefont {R.~E.}\ \bibnamefont {Parks}}, \
		and\ \bibinfo {author} {\bibfnamefont {A.~M.}\ \bibnamefont {Sleeper}},\
	}\href@noop {} {\bibfield  {journal} {\bibinfo  {journal} {Phys. Rev. Lett.}\
		}\textbf {\bibinfo {volume} {14}},\ \bibinfo {pages} {1029} (\bibinfo {year}
		{1965})}\BibitemShut {NoStop}%
	\bibitem [{\citenamefont {Bloembergen}\ \emph {et~al.}(1968)\citenamefont
		{Bloembergen}, \citenamefont {Chang}, \citenamefont {Jha},\ and\
		\citenamefont {Lee}}]{Bloembergen1968}%
	\BibitemOpen
	\bibfield  {author} {\bibinfo {author} {\bibfnamefont {N.}~\bibnamefont
			{Bloembergen}}, \bibinfo {author} {\bibfnamefont {R.~K.}\ \bibnamefont
			{Chang}}, \bibinfo {author} {\bibfnamefont {S.~S.}\ \bibnamefont {Jha}}, \
		and\ \bibinfo {author} {\bibfnamefont {C.~H.}\ \bibnamefont {Lee}},\
	}\href@noop {} {\bibfield  {journal} {\bibinfo  {journal} {Phys. Rev.}\
		}\textbf {\bibinfo {volume} {174}},\ \bibinfo {pages} {813} (\bibinfo {year}
		{1968})}\BibitemShut {NoStop}%
	\bibitem [{\citenamefont {Simon}\ \emph {et~al.}(1974)\citenamefont {Simon},
		\citenamefont {Mitchell},\ and\ \citenamefont {Watson}}]{Simon1974}%
	\BibitemOpen
	\bibfield  {author} {\bibinfo {author} {\bibfnamefont {H.~J.}\ \bibnamefont
			{Simon}}, \bibinfo {author} {\bibfnamefont {D.~E.}\ \bibnamefont {Mitchell}},
		\ and\ \bibinfo {author} {\bibfnamefont {J.~G.}\ \bibnamefont {Watson}},\
	}\href@noop {} {\bibfield  {journal} {\bibinfo  {journal} {Phys. Rev. Lett.}\
		}\textbf {\bibinfo {volume} {33}},\ \bibinfo {pages} {1531} (\bibinfo {year}
		{1974})}\BibitemShut {NoStop}%
	\bibitem [{\citenamefont {M\"{a}kitalo}\ \emph {et~al.}(2011)\citenamefont
		{M\"{a}kitalo}, \citenamefont {Suuriniemi},\ and\ \citenamefont
		{Kauranen}}]{Makitalo2011}%
	\BibitemOpen
	\bibfield  {author} {\bibinfo {author} {\bibfnamefont {J.}~\bibnamefont
			{M\"{a}kitalo}}, \bibinfo {author} {\bibfnamefont {S.}~\bibnamefont
			{Suuriniemi}}, \ and\ \bibinfo {author} {\bibfnamefont {M.}~\bibnamefont
			{Kauranen}},\ }\href@noop {} {\bibfield  {journal} {\bibinfo  {journal} {Opt.
				Express}\ }\textbf {\bibinfo {volume} {19}},\ \bibinfo {pages} {23386}
		(\bibinfo {year} {2011})}\BibitemShut {NoStop}%
	\bibitem [{\citenamefont {Zhang}\ \emph {et~al.}(2011)\citenamefont {Zhang},
		\citenamefont {Grady}, \citenamefont {Ayala-Orozco},\ and\ \citenamefont
		{Halas}}]{Zhang2011}%
	\BibitemOpen
	\bibfield  {author} {\bibinfo {author} {\bibfnamefont {Y.}~\bibnamefont
			{Zhang}}, \bibinfo {author} {\bibfnamefont {N.~K.}\ \bibnamefont {Grady}},
		\bibinfo {author} {\bibfnamefont {C.}~\bibnamefont {Ayala-Orozco}}, \ and\
		\bibinfo {author} {\bibfnamefont {N.~J.}\ \bibnamefont {Halas}},\ }\href@noop
	{} {\bibfield  {journal} {\bibinfo  {journal} {Nano Letters}\ }\textbf
		{\bibinfo {volume} {11}},\ \bibinfo {pages} {5519} (\bibinfo {year}
		{2011})}\BibitemShut {NoStop}%
	\bibitem [{\citenamefont {Konishi}\ \emph {et~al.}(2014)\citenamefont
		{Konishi}, \citenamefont {Higuchi}, \citenamefont {Li}, \citenamefont
		{Larsson}, \citenamefont {Ishii},\ and\ \citenamefont
		{Kuwata-Gonokami}}]{Konishi2014}%
	\BibitemOpen
	\bibfield  {author} {\bibinfo {author} {\bibfnamefont {K.}~\bibnamefont
			{Konishi}}, \bibinfo {author} {\bibfnamefont {T.}~\bibnamefont {Higuchi}},
		\bibinfo {author} {\bibfnamefont {J.}~\bibnamefont {Li}}, \bibinfo {author}
		{\bibfnamefont {J.}~\bibnamefont {Larsson}}, \bibinfo {author} {\bibfnamefont
			{S.}~\bibnamefont {Ishii}}, \ and\ \bibinfo {author} {\bibfnamefont
			{M.}~\bibnamefont {Kuwata-Gonokami}},\ }\href@noop {} {\bibfield  {journal}
		{\bibinfo  {journal} {Phys. Rev. Lett.}\ }\textbf {\bibinfo {volume} {112}},\
		\bibinfo {pages} {135502} (\bibinfo {year} {2014})}\BibitemShut {NoStop}%
	\bibitem [{\citenamefont {O'Brien}\ \emph {et~al.}(2015)\citenamefont
		{O'Brien}, \citenamefont {Suchowski}, \citenamefont {Rho}, \citenamefont
		{Salandrino}, \citenamefont {Kante}, \citenamefont {Yin},\ and\ \citenamefont
		{Zhang}}]{OBrien2015}%
	\BibitemOpen
	\bibfield  {author} {\bibinfo {author} {\bibfnamefont {K.}~\bibnamefont
			{O'Brien}}, \bibinfo {author} {\bibfnamefont {H.}~\bibnamefont {Suchowski}},
		\bibinfo {author} {\bibfnamefont {J.}~\bibnamefont {Rho}}, \bibinfo {author}
		{\bibfnamefont {A.}~\bibnamefont {Salandrino}}, \bibinfo {author}
		{\bibfnamefont {B.}~\bibnamefont {Kante}}, \bibinfo {author} {\bibfnamefont
			{X.}~\bibnamefont {Yin}}, \ and\ \bibinfo {author} {\bibfnamefont
			{X.}~\bibnamefont {Zhang}},\ }\href@noop {} {\bibfield  {journal} {\bibinfo
			{journal} {Nature Materials}\ }\textbf {\bibinfo {volume} {14}},\ \bibinfo
		{pages} {379} (\bibinfo {year} {2015})}\BibitemShut {NoStop}%
	\bibitem [{\citenamefont {Czaplicki}\ \emph {et~al.}(2015)\citenamefont
		{Czaplicki}, \citenamefont {M\"{a}kitalo}, \citenamefont {Siikanen},
		\citenamefont {Husu}, \citenamefont {Lehtolahti}, \citenamefont {Kuittinen},\
		and\ \citenamefont {Kauranen}}]{Czaplicki2015}%
	\BibitemOpen
	\bibfield  {author} {\bibinfo {author} {\bibfnamefont {R.}~\bibnamefont
			{Czaplicki}}, \bibinfo {author} {\bibfnamefont {J.}~\bibnamefont
			{M\"{a}kitalo}}, \bibinfo {author} {\bibfnamefont {R.}~\bibnamefont
			{Siikanen}}, \bibinfo {author} {\bibfnamefont {H.}~\bibnamefont {Husu}},
		\bibinfo {author} {\bibfnamefont {J.}~\bibnamefont {Lehtolahti}}, \bibinfo
		{author} {\bibfnamefont {M.}~\bibnamefont {Kuittinen}}, \ and\ \bibinfo
		{author} {\bibfnamefont {M.}~\bibnamefont {Kauranen}},\ }\href@noop {}
	{\bibfield  {journal} {\bibinfo  {journal} {Nano Letters}\ }\textbf {\bibinfo
			{volume} {15}},\ \bibinfo {pages} {530} (\bibinfo {year} {2015})}\BibitemShut
	{NoStop}%
	\bibitem [{\citenamefont {Celebrano}\ \emph {et~al.}(2015)\citenamefont
		{Celebrano}, \citenamefont {Wu}, \citenamefont {Baselli}, \citenamefont
		{Gro{\ss}mann}, \citenamefont {Biagioni}, \citenamefont {Locatelli},
		\citenamefont {De~Angelis}, \citenamefont {Cerullo}, \citenamefont
		{Osellame}, \citenamefont {Hecht}, \citenamefont {Du{\`o}}, \citenamefont
		{Ciccacci},\ and\ \citenamefont {Finazzi}}]{Celebrano2015}%
	\BibitemOpen
	\bibfield  {author} {\bibinfo {author} {\bibfnamefont {M.}~\bibnamefont
			{Celebrano}}, \bibinfo {author} {\bibfnamefont {X.}~\bibnamefont {Wu}},
		\bibinfo {author} {\bibfnamefont {M.}~\bibnamefont {Baselli}}, \bibinfo
		{author} {\bibfnamefont {S.}~\bibnamefont {Gro{\ss}mann}}, \bibinfo {author}
		{\bibfnamefont {P.}~\bibnamefont {Biagioni}}, \bibinfo {author}
		{\bibfnamefont {A.}~\bibnamefont {Locatelli}}, \bibinfo {author}
		{\bibfnamefont {C.}~\bibnamefont {De~Angelis}}, \bibinfo {author}
		{\bibfnamefont {G.}~\bibnamefont {Cerullo}}, \bibinfo {author} {\bibfnamefont
			{R.}~\bibnamefont {Osellame}}, \bibinfo {author} {\bibfnamefont
			{B.}~\bibnamefont {Hecht}}, \bibinfo {author} {\bibfnamefont
			{L.}~\bibnamefont {Du{\`o}}}, \bibinfo {author} {\bibfnamefont
			{F.}~\bibnamefont {Ciccacci}}, \ and\ \bibinfo {author} {\bibfnamefont
			{M.}~\bibnamefont {Finazzi}},\ }\href@noop {} {\bibfield  {journal} {\bibinfo
			{journal} {Nature Nanotechnology}\ }\textbf {\bibinfo {volume} {10}},\
		\bibinfo {pages} {412} (\bibinfo {year} {2015})}\BibitemShut {NoStop}%
	\bibitem [{\citenamefont {Gennaro}\ \emph {et~al.}(2016)\citenamefont
		{Gennaro}, \citenamefont {Rahmani}, \citenamefont {Giannini}, \citenamefont
		{Aouani}, \citenamefont {Sidiropoulos}, \citenamefont {Navarro-Cía},
		\citenamefont {Maier},\ and\ \citenamefont {Oulton}}]{Gennaro2016}%
	\BibitemOpen
	\bibfield  {author} {\bibinfo {author} {\bibfnamefont {S.~D.}\ \bibnamefont
			{Gennaro}}, \bibinfo {author} {\bibfnamefont {M.}~\bibnamefont {Rahmani}},
		\bibinfo {author} {\bibfnamefont {V.}~\bibnamefont {Giannini}}, \bibinfo
		{author} {\bibfnamefont {H.}~\bibnamefont {Aouani}}, \bibinfo {author}
		{\bibfnamefont {T.~P.~H.}\ \bibnamefont {Sidiropoulos}}, \bibinfo {author}
		{\bibfnamefont {M.}~\bibnamefont {Navarro-Cía}}, \bibinfo {author}
		{\bibfnamefont {S.~A.}\ \bibnamefont {Maier}}, \ and\ \bibinfo {author}
		{\bibfnamefont {R.~F.}\ \bibnamefont {Oulton}},\ }\href@noop {} {\bibfield
		{journal} {\bibinfo  {journal} {Nano Letters}\ }\textbf {\bibinfo {volume}
			{16}},\ \bibinfo {pages} {5278} (\bibinfo {year} {2016})}\BibitemShut
	{NoStop}%
	\bibitem [{\citenamefont {G\'{o}mez-Tornero}\ \emph {et~al.}(2017)\citenamefont
		{G\'{o}mez-Tornero}, \citenamefont {Tserkezis}, \citenamefont {Mateos},
		\citenamefont {Baus\'{a}},\ and\ \citenamefont
		{Ram\'{i}rez}}]{Gomez-Tornero2017}%
	\BibitemOpen
	\bibfield  {author} {\bibinfo {author} {\bibfnamefont {A.}~\bibnamefont
			{G\'{o}mez-Tornero}}, \bibinfo {author} {\bibfnamefont {C.}~\bibnamefont
			{Tserkezis}}, \bibinfo {author} {\bibfnamefont {L.}~\bibnamefont {Mateos}},
		\bibinfo {author} {\bibfnamefont {L.~E.}\ \bibnamefont {Baus\'{a}}}, \ and\
		\bibinfo {author} {\bibfnamefont {M.~O.}\ \bibnamefont {Ram\'{i}rez}},\
	}\href@noop {} {\bibfield  {journal} {\bibinfo  {journal} {Advanced
				Materials}\ }\textbf {\bibinfo {volume} {29}},\ \bibinfo {pages} {1605267}
		(\bibinfo {year} {2017})}\BibitemShut {NoStop}%
	\bibitem [{\citenamefont {Yang}\ \emph {et~al.}(2017)\citenamefont {Yang},
		\citenamefont {Butet}, \citenamefont {Yan}, \citenamefont {Bernasconi},\ and\
		\citenamefont {Martin}}]{Yang2017}%
	\BibitemOpen
	\bibfield  {author} {\bibinfo {author} {\bibfnamefont {K.-Y.}\ \bibnamefont
			{Yang}}, \bibinfo {author} {\bibfnamefont {J.}~\bibnamefont {Butet}},
		\bibinfo {author} {\bibfnamefont {C.}~\bibnamefont {Yan}}, \bibinfo {author}
		{\bibfnamefont {G.~D.}\ \bibnamefont {Bernasconi}}, \ and\ \bibinfo {author}
		{\bibfnamefont {O.~J.~F.}\ \bibnamefont {Martin}},\ }\href@noop {} {\bibfield
		{journal} {\bibinfo  {journal} {ACS Photonics}\ }\textbf {\bibinfo {volume}
			{4}},\ \bibinfo {pages} {1522} (\bibinfo {year} {2017})}\BibitemShut
	{NoStop}%
	\bibitem [{\citenamefont {Chervinskii}\ \emph {et~al.}(2018)\citenamefont
		{Chervinskii}, \citenamefont {Koskinen}, \citenamefont {Scherbak},
		\citenamefont {Kauranen},\ and\ \citenamefont {Lipovskii}}]{Chervinskii2018}%
	\BibitemOpen
	\bibfield  {author} {\bibinfo {author} {\bibfnamefont {S.}~\bibnamefont
			{Chervinskii}}, \bibinfo {author} {\bibfnamefont {K.}~\bibnamefont
			{Koskinen}}, \bibinfo {author} {\bibfnamefont {S.}~\bibnamefont {Scherbak}},
		\bibinfo {author} {\bibfnamefont {M.}~\bibnamefont {Kauranen}}, \ and\
		\bibinfo {author} {\bibfnamefont {A.}~\bibnamefont {Lipovskii}},\ }\href@noop
	{} {\bibfield  {journal} {\bibinfo  {journal} {Phys. Rev. Lett.}\ }\textbf
		{\bibinfo {volume} {120}},\ \bibinfo {pages} {113902} (\bibinfo {year}
		{2018})}\BibitemShut {NoStop}%
	\bibitem [{\citenamefont {Chen}\ \emph {et~al.}(1979)\citenamefont {Chen},
		\citenamefont {de~Castro},\ and\ \citenamefont {Shen}}]{Chen1979}%
	\BibitemOpen
	\bibfield  {author} {\bibinfo {author} {\bibfnamefont {C.~K.}\ \bibnamefont
			{Chen}}, \bibinfo {author} {\bibfnamefont {A.~R.~B.}\ \bibnamefont
			{de~Castro}}, \ and\ \bibinfo {author} {\bibfnamefont {Y.~R.}\ \bibnamefont
			{Shen}},\ }\href@noop {} {\bibfield  {journal} {\bibinfo  {journal} {Opt.
				Lett.}\ }\textbf {\bibinfo {volume} {4}},\ \bibinfo {pages} {393} (\bibinfo
		{year} {1979})}\BibitemShut {NoStop}%
	\bibitem [{\citenamefont {Viarbitskaya}\ \emph {et~al.}(2015)\citenamefont
		{Viarbitskaya}, \citenamefont {Demichel}, \citenamefont {Cluzel},
		\citenamefont {Colas~des Francs},\ and\ \citenamefont
		{Bouhelier}}]{Viarbitskaya2015}%
	\BibitemOpen
	\bibfield  {author} {\bibinfo {author} {\bibfnamefont {S.}~\bibnamefont
			{Viarbitskaya}}, \bibinfo {author} {\bibfnamefont {O.}~\bibnamefont
			{Demichel}}, \bibinfo {author} {\bibfnamefont {B.}~\bibnamefont {Cluzel}},
		\bibinfo {author} {\bibfnamefont {G.}~\bibnamefont {Colas~des Francs}}, \
		and\ \bibinfo {author} {\bibfnamefont {A.}~\bibnamefont {Bouhelier}},\
	}\href@noop {} {\bibfield  {journal} {\bibinfo  {journal} {Phys. Rev. Lett.}\
		}\textbf {\bibinfo {volume} {115}},\ \bibinfo {pages} {197401} (\bibinfo
		{year} {2015})}\BibitemShut {NoStop}%
	\bibitem [{\citenamefont {de~Hoogh}\ \emph {et~al.}(2016)\citenamefont
		{de~Hoogh}, \citenamefont {Opheij}, \citenamefont {Wulf}, \citenamefont
		{Rotenberg},\ and\ \citenamefont {Kuipers}}]{deHoogh2016}%
	\BibitemOpen
	\bibfield  {author} {\bibinfo {author} {\bibfnamefont {A.}~\bibnamefont
			{de~Hoogh}}, \bibinfo {author} {\bibfnamefont {A.}~\bibnamefont {Opheij}},
		\bibinfo {author} {\bibfnamefont {M.}~\bibnamefont {Wulf}}, \bibinfo {author}
		{\bibfnamefont {N.}~\bibnamefont {Rotenberg}}, \ and\ \bibinfo {author}
		{\bibfnamefont {L.}~\bibnamefont {Kuipers}},\ }\href@noop {} {\bibfield
		{journal} {\bibinfo  {journal} {ACS Photonics}\ }\textbf {\bibinfo {volume}
			{3}},\ \bibinfo {pages} {1446} (\bibinfo {year} {2016})}\BibitemShut
	{NoStop}%
	\bibitem [{\citenamefont {Li}\ \emph {et~al.}(2017)\citenamefont {Li},
		\citenamefont {Kang}, \citenamefont {Shi}, \citenamefont {Wu}, \citenamefont
		{Zhang},\ and\ \citenamefont {Xu}}]{Li2017}%
	\BibitemOpen
	\bibfield  {author} {\bibinfo {author} {\bibfnamefont {Y.}~\bibnamefont
			{Li}}, \bibinfo {author} {\bibfnamefont {M.}~\bibnamefont {Kang}}, \bibinfo
		{author} {\bibfnamefont {J.}~\bibnamefont {Shi}}, \bibinfo {author}
		{\bibfnamefont {K.}~\bibnamefont {Wu}}, \bibinfo {author} {\bibfnamefont
			{S.}~\bibnamefont {Zhang}}, \ and\ \bibinfo {author} {\bibfnamefont
			{H.}~\bibnamefont {Xu}},\ }\href@noop {} {\bibfield  {journal} {\bibinfo
			{journal} {Nano Letters}\ }\textbf {\bibinfo {volume} {17}},\ \bibinfo
		{pages} {7803} (\bibinfo {year} {2017})}\BibitemShut {NoStop}%
	\bibitem [{\citenamefont {Lan}\ \emph {et~al.}(2015)\citenamefont {Lan},
		\citenamefont {Kang}, \citenamefont {Schoen}, \citenamefont {Rodrigues},
		\citenamefont {Cui}, \citenamefont {Brongersma},\ and\ \citenamefont
		{Cai}}]{Lan:2015bi}%
	\BibitemOpen
	\bibfield  {author} {\bibinfo {author} {\bibfnamefont {S.}~\bibnamefont
			{Lan}}, \bibinfo {author} {\bibfnamefont {L.}~\bibnamefont {Kang}}, \bibinfo
		{author} {\bibfnamefont {D.~T.}\ \bibnamefont {Schoen}}, \bibinfo {author}
		{\bibfnamefont {S.~P.}\ \bibnamefont {Rodrigues}}, \bibinfo {author}
		{\bibfnamefont {Y.}~\bibnamefont {Cui}}, \bibinfo {author} {\bibfnamefont
			{M.~L.}\ \bibnamefont {Brongersma}}, \ and\ \bibinfo {author} {\bibfnamefont
			{W.}~\bibnamefont {Cai}},\ }\href@noop {} {\bibfield  {journal} {\bibinfo
			{journal} {Nature Materials}\ }\textbf {\bibinfo {volume} {14}},\ \bibinfo
		{pages} {807} (\bibinfo {year} {2015})}\BibitemShut {NoStop}%
	\bibitem [{\citenamefont {Geisler}\ \emph {et~al.}(2013)\citenamefont
		{Geisler}, \citenamefont {Razinskas}, \citenamefont {Krauss}, \citenamefont
		{Wu}, \citenamefont {Rewitz}, \citenamefont {Tuchscherer}, \citenamefont
		{Goetz}, \citenamefont {Huang}, \citenamefont {Brixner},\ and\ \citenamefont
		{Hecht}}]{Geisler2013}%
	\BibitemOpen
	\bibfield  {author} {\bibinfo {author} {\bibfnamefont {P.}~\bibnamefont
			{Geisler}}, \bibinfo {author} {\bibfnamefont {G.}~\bibnamefont {Razinskas}},
		\bibinfo {author} {\bibfnamefont {E.}~\bibnamefont {Krauss}}, \bibinfo
		{author} {\bibfnamefont {X.-F.}\ \bibnamefont {Wu}}, \bibinfo {author}
		{\bibfnamefont {C.}~\bibnamefont {Rewitz}}, \bibinfo {author} {\bibfnamefont
			{P.}~\bibnamefont {Tuchscherer}}, \bibinfo {author} {\bibfnamefont
			{S.}~\bibnamefont {Goetz}}, \bibinfo {author} {\bibfnamefont {C.-B.}\
			\bibnamefont {Huang}}, \bibinfo {author} {\bibfnamefont {T.}~\bibnamefont
			{Brixner}}, \ and\ \bibinfo {author} {\bibfnamefont {B.}~\bibnamefont
			{Hecht}},\ }\href@noop {} {\bibfield  {journal} {\bibinfo  {journal} {Phys.
				Rev. Lett.}\ }\textbf {\bibinfo {volume} {111}},\ \bibinfo {pages} {183901}
		(\bibinfo {year} {2013})}\BibitemShut {NoStop}%
	\bibitem [{\citenamefont {Wu}\ \emph {et~al.}(2017)\citenamefont {Wu},
		\citenamefont {Jiang}, \citenamefont {Razinskas}, \citenamefont {Huo},
		\citenamefont {Zhang}, \citenamefont {Kamp}, \citenamefont {Rastelli},
		\citenamefont {Schmidt}, \citenamefont {Hecht}, \citenamefont {Lindfors},\
		and\ \citenamefont {Lippitz}}]{Wu2017}%
	\BibitemOpen
	\bibfield  {author} {\bibinfo {author} {\bibfnamefont {X.}~\bibnamefont
			{Wu}}, \bibinfo {author} {\bibfnamefont {P.}~\bibnamefont {Jiang}}, \bibinfo
		{author} {\bibfnamefont {G.}~\bibnamefont {Razinskas}}, \bibinfo {author}
		{\bibfnamefont {Y.}~\bibnamefont {Huo}}, \bibinfo {author} {\bibfnamefont
			{H.}~\bibnamefont {Zhang}}, \bibinfo {author} {\bibfnamefont
			{M.}~\bibnamefont {Kamp}}, \bibinfo {author} {\bibfnamefont {A.}~\bibnamefont
			{Rastelli}}, \bibinfo {author} {\bibfnamefont {O.~G.}\ \bibnamefont
			{Schmidt}}, \bibinfo {author} {\bibfnamefont {B.}~\bibnamefont {Hecht}},
		\bibinfo {author} {\bibfnamefont {K.}~\bibnamefont {Lindfors}}, \ and\
		\bibinfo {author} {\bibfnamefont {M.}~\bibnamefont {Lippitz}},\ }\href@noop
	{} {\bibfield  {journal} {\bibinfo  {journal} {Nano Letters}\ }\textbf
		{\bibinfo {volume} {17}},\ \bibinfo {pages} {4291} (\bibinfo {year}
		{2017})}\BibitemShut {NoStop}%
	\bibitem [{\citenamefont {Dai}\ \emph {et~al.}(2014)\citenamefont {Dai},
		\citenamefont {Lin}, \citenamefont {Huang},\ and\ \citenamefont
		{Huang}}]{Dai2014}%
	\BibitemOpen
	\bibfield  {author} {\bibinfo {author} {\bibfnamefont {W.-H.}\ \bibnamefont
			{Dai}}, \bibinfo {author} {\bibfnamefont {F.-C.}\ \bibnamefont {Lin}},
		\bibinfo {author} {\bibfnamefont {C.-B.}\ \bibnamefont {Huang}}, \ and\
		\bibinfo {author} {\bibfnamefont {J.-S.}\ \bibnamefont {Huang}},\ }\href@noop
	{} {\bibfield  {journal} {\bibinfo  {journal} {Nano Letters}\ }\textbf
		{\bibinfo {volume} {14}},\ \bibinfo {pages} {3881} (\bibinfo {year}
		{2014})}\BibitemShut {NoStop}%
	\bibitem [{\citenamefont {Roke}\ \emph {et~al.}(2004)\citenamefont {Roke},
		\citenamefont {Bonn},\ and\ \citenamefont {Petukhov}}]{Roke:2004cf}%
	\BibitemOpen
	\bibfield  {author} {\bibinfo {author} {\bibfnamefont {S.}~\bibnamefont
			{Roke}}, \bibinfo {author} {\bibfnamefont {M.}~\bibnamefont {Bonn}}, \ and\
		\bibinfo {author} {\bibfnamefont {A.~V.}\ \bibnamefont {Petukhov}},\
	}\href@noop {} {\bibfield  {journal} {\bibinfo  {journal} {Physical Review
				B}\ }\textbf {\bibinfo {volume} {70}},\ \bibinfo {pages} {9374} (\bibinfo
		{year} {2004})}\BibitemShut {NoStop}%
	\bibitem [{\citenamefont {Rewitz}\ \emph {et~al.}(2014)\citenamefont {Rewitz},
		\citenamefont {Razinskas}, \citenamefont {Geisler}, \citenamefont {Krauss},
		\citenamefont {Goetz}, \citenamefont {Paw\l{}owska}, \citenamefont {Hecht},\
		and\ \citenamefont {Brixner}}]{Rewitz2014}%
	\BibitemOpen
	\bibfield  {author} {\bibinfo {author} {\bibfnamefont {C.}~\bibnamefont
			{Rewitz}}, \bibinfo {author} {\bibfnamefont {G.}~\bibnamefont {Razinskas}},
		\bibinfo {author} {\bibfnamefont {P.}~\bibnamefont {Geisler}}, \bibinfo
		{author} {\bibfnamefont {E.}~\bibnamefont {Krauss}}, \bibinfo {author}
		{\bibfnamefont {S.}~\bibnamefont {Goetz}}, \bibinfo {author} {\bibfnamefont
			{M.}~\bibnamefont {Paw\l{}owska}}, \bibinfo {author} {\bibfnamefont
			{B.}~\bibnamefont {Hecht}}, \ and\ \bibinfo {author} {\bibfnamefont
			{T.}~\bibnamefont {Brixner}},\ }\href@noop {} {\bibfield  {journal} {\bibinfo
			{journal} {Phys. Rev. Applied}\ }\textbf {\bibinfo {volume} {1}},\ \bibinfo
		{pages} {014007} (\bibinfo {year} {2014})}\BibitemShut {NoStop}%
\end{thebibliography}
\end{document}



\title{Supplementary material to 'Symmetry-controlled second-harmonic generation in a plasmonic waveguide'}

\author{Tzu-Yu Chen}
\affiliation
{Institute of Photonics Technologies, National Tsing Hua University, Hsinchu 30013, Taiwan}
\affiliation{International Intercollegiate PhD Program, National Tsing Hua University, Hsinchu 30013, Taiwan}

\author{Julian Obermeier}
\affiliation{Department of Physics, University of Bayreuth, 95440 Bayreuth, Germany}

\author{Thorsten Schumacher}
\affiliation{Department of Physics, University of Bayreuth, 95440 Bayreuth, Germany}

\author{Fan-Cheng Lin}
\affiliation{Department of Chemistry, National Tsing Hua University, Hsinchu 30013, Taiwan}  

\author{Jer-Shing Huang}
\affiliation{Leibniz Institute of Photonic Technology, 07745 Jena, Germany}
\affiliation{Department of Electrophysics, National Chiao Tung University, Hsinchu 30010, Taiwan}
\affiliation{Research Center for Applied Sciences, Academia Sinica, Taipei 115-29, Taiwan}

\author{Markus Lippitz}
\email{markus.lippitz@uni-bayreuth.de}
\affiliation{Department of Physics, University of Bayreuth, 95440 Bayreuth, Germany}

\author{Chen-Bin Huang}
\email{robin@ee.nthu.edu.tw}
\affiliation
{Institute of Photonics Technologies, National Tsing Hua University, Hsinchu 30013, Taiwan}
\affiliation{International Intercollegiate PhD Program, National Tsing Hua University, Hsinchu 30013, Taiwan}
\affiliation{Research Center for Applied Sciences, Academia Sinica, Taipei 115-29, Taiwan}


\maketitle


\section{Device fabrication and experimental setup}

\begin{figure}
	\includegraphics[width=\columnwidth]{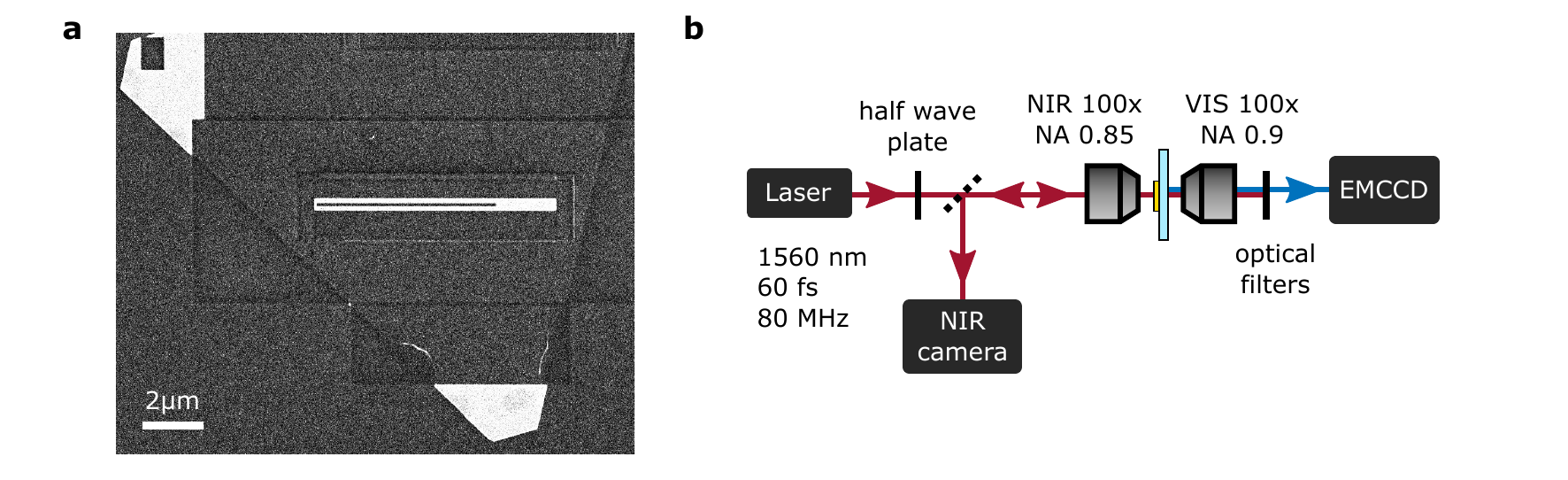}
	\caption{\textbf{Fabricated TWTL and experimental setup.} (a) SEM image of a TWTL fabricated by focused ion-beam milling. For the top-down fabrication procedure single crystalin gold flakes are used. (b) The experimental setup is a home-built dual-confocal microscope that allows imaging of the fundamental and second harmonic signals.  
		\label{fig:FIBsetup}}
\end{figure}

The plasmonic TWTL is fabricated by focused ion-beam milling (FEI Helios) out of a single crystalline gold flake with a height of 60~nm (Fig.~\ref{fig:FIBsetup}a). The flake is deposited on a glass substrate that is coated with a conductive indium tin oxide (ITO) layer of 40~nm. The width of both wires is set to 140~nm, while the gap separation is 100~nm. The linked input antenna has a width of 80~nm and is designed to allow the efficient excitation of both fundamental modes.
A home-built dual-confocal transmission microscope is used to perform all measurements (Fig.~\ref{fig:FIBsetup}b). The excitation laser is a passively mode-locked fiber laser centered at 1560~nm, producing 60~fs transform-limited optical pulses at 80~MHz repetition rate with a maximum average power of 100~mW (Menlo Systems T-Light). The linearly polarized laser pulses are focused by a NIR long working distance 100x objective lens with NA=0.85 (Olympus LCPLN100XIR) to excite the plasmonic TWTL from the substrate side. Fundamental harmonic images are recorded in reflection by a NIR camera (Xenics Xeva-1.7-320). 
The SH signals are collected in transmission from the air side by a visible 100x objective with NA=0.9 (Olympus MPLFLN100X) and recorded by an electron-multiplied charge-coupled device (EMCCD, Andor iXon 897U-CS0-EXF). A short pass filter (Thorlabs FESH0900) and a band-pass filter (Thorlabs FL780-10) ensure that the recorded images show purely SH signals.

\section{Numerical simulation}
To calculate the 2D modes of our plasmonic waveguide we use the commercial FEM software Comsol Multiphysics. Each of the two wires has a width of 140~nm and a height of 60~nm (edge rounding 25~nm, gap 100~nm). The optical properties of the waveguide consisting of gold are taken as interpolated data from Johnson and Christy \cite{JC:1972}. The homogeneous environment has the refractive index n=1. The environment was chosen homogeneous for a simplified representation but our argumentation is not affected or changed by adding a substrate.
The mode analysis solves for the full vectorial fields. Since there are no resonances in the cross section of the TWTL the spatial shape of the modes is nearly independent of the wavelength. The subsequent calculation of the nonlinear polarization is performed in Matlab.

\section{Full vectorial model}
In general, the second order nonlinear polarization is defined as
\begin{eqnarray}
P^{(2)}_i(2\omega) &=& \epsilon_0 \sum\limits_{jk} \chi_{ijk}^{(2)} E_j(\omega)E_k(\omega) \quad \text{with} \quad i,j,k \in \{x,y,z\} \quad .
\label{eq:e1}
\end{eqnarray}
with $\epsilon_0$ being the vacuum permittivity, $\chi_{ijk}^{(2)}$ the second order susceptibility, and $E_{j,k}(\omega) $ the electric field components at frequency $\omega$.
However, due to the centro-symmetry of gold, the bulk contribution of the material vanishes and only surface nonlinearities have to be considered. Here, the $\chi^{(2)}$-tensor reduces to seven nonvanishing elements in a local coordinate system, at the surface of the structure under investigation \cite{Makitalo2011}. The transformation $(x,y,z) \rightarrow (t,n,s)$ is described by the rotation matrix 
\begin{eqnarray}
\underline{\underline{D}} &=& \begin{pmatrix}
\hat{\text{e}}_t \cdot \hat{\text{e}}_x &  \hat{\text{e}}_t \cdot \hat{\text{e}}_y & 0\\
\hat{\text{e}}_n \cdot \hat{\text{e}}_x &  \hat{\text{e}}_n \cdot \hat{\text{e}}_y & 0\\
0 & 0 & 1		\end{pmatrix} =
\begin{pmatrix}
n_y & -n_x & 0 \\
n_x & n_y  & 0 \\
0 & 0    & 1\end{pmatrix} \quad ,
\end{eqnarray}

where $n_x$ and $n_y$ are the vector components of the surface normal.
The nonvanishing $\chi^{(2)}$ components are $\chi^{(2)}_1 \equiv \chi^{(2)}_{nnn}$, $\chi^{(2)}_2 \equiv \chi^{(2)}_{nss} = \chi^{(2)}_{ntt}$ and $\chi^{(2)}_3 \equiv \chi^{(2)}_{ssn} = \chi^{(2)}_{sns} = \chi^{(2)}_{ttn} = \chi^{(2)}_{tnt}$. After a local evaluation of $P^{(2)}$ at the surface and a back-transformation to Cartesian coordinates we obtain
\begin{eqnarray}
P^{(2)}_x &=& \epsilon_0 n_x \chi^{(2)}_1 (E_x n_x + E_y n_y)^2 \nonumber\\
& & +\,\epsilon_0 n_x \chi^{(2)}_2 \left[ E_z^2 + (E_x n_y - E_y n_x)^2\right] \nonumber\\
& & +\,2 \epsilon_0 n_y \chi^{(2)}_3 (E_x n_x + E_y n_y) (E_x n_y - E_y n_x) \\
P^{(2)}_y &=& \epsilon_0 n_y \chi^{(2)}_1 (E_x n_x + E_y n_y)^2 \nonumber\\
& & +\,\epsilon_0 n_y \chi^{(2)}_2 \left[ E_z^2 + (E_x n_y - E_y n_x)^2\right] \nonumber\\
& & -\,2 \epsilon_0 n_x \chi^{(2)}_3 (E_x n_x + E_y n_y) (E_x n_y - E_y n_x) \\
P^{(2)}_z &=& 2 \epsilon_0 \chi^{(2)}_3 E_z (E_x n_x + E_y n_y)
\label{eq:e6}
\end{eqnarray}
Furthermore, we define the coupling efficiency $\eta$ of the nonlinear polarization $\mathbf{P}^{(2)}(2 \omega)$ to the waveguide modes $\mathbf{E}(2 \omega)$ similar to Ref. \cite{OBrien2015} as
\begin{equation}
\eta = \left| \int_{\partial A} \mathbf{P}^{(2)}(2 \omega)\, \cdot \,  \mathbf{E}(2 \omega)^* \, ds \right |^2 \quad ,
\label{eq:eta}
\end{equation}
with $\partial A$ being the path along the 1D surface of the structure's cross section (see also main text). \newline
\begin{figure}
	\includegraphics[width=\columnwidth]{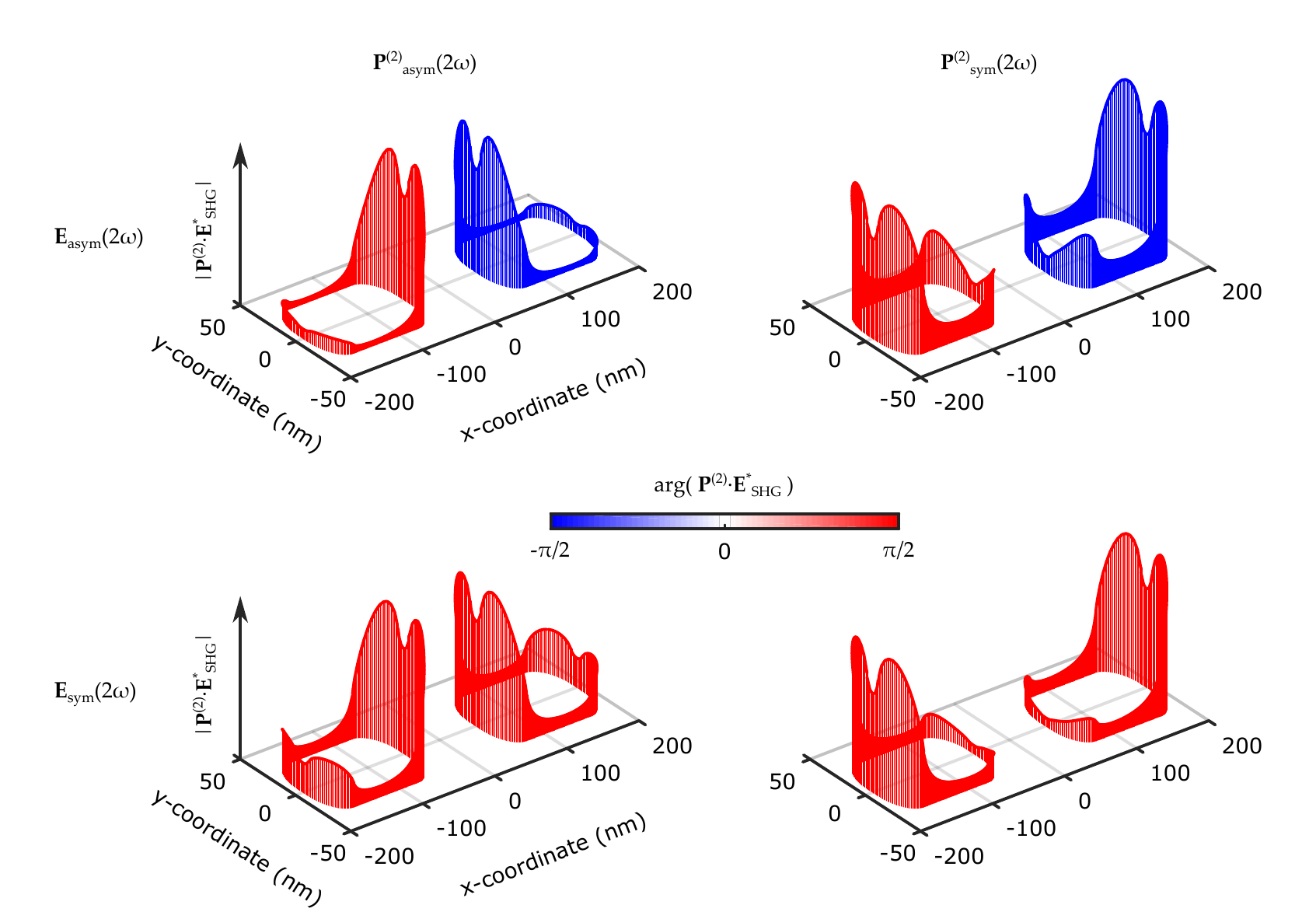}
	\caption{\textbf{Spatially resolved complex integrands of the four coupling integrals.} In all graphs, the x-y-plane shows the  cross section through the waveguide, the z-axis the integrand as amplitude and color-coded phase around the structure's surface. The amplitude of all four plots is normalized to the individual maximum since absolute magnitudes are not relevant for our  purely symmetry based argumentation.
		\label{fig:overlap}}
\end{figure}
In the following we compute the coupling efficiency from the fundamental into the second harmonic waveguide modes, applying the full vectorial model. As given by Ref. \cite{Kauranen:2009}, we use the relative magnitudes $\chi^{(2)}_{1}=250$, $\chi^{(2)}_{2}=1$ and $\chi^{(2)}_{3}=3.6$ for the nonlinear susceptibility. With the nonlinear polarizations of the two fundamental modes and the two available second harmonic modes, we obtain four independent overlap integrals (eq.~\ref{eq:eta}). For visualization, the integrands of these integrals are shown in Fig.~\ref{fig:overlap}. It covers all four combination where the columns give the two nonlinear polarizations, the two rows the waveguide modes at the second harmonic frequency. In each plot, the x-y-plane shows the cross section through the waveguide and the z-direction the absolute value $|\mathbf{P}^{(2)}(2 \omega)\, \cdot \,  \mathbf{E}(2 \omega)^*|$ of the integrand, evaluated at the surface of the structure. In each case, the amplitudes at the left and right waveguide branch show the same symmetry as the structure itself. Furthermore, the phase of the integrand is continuously color-coded from $-\pi/2$ (blue) over $0$ (white) to $\pi/2$ (red). We find, that either both branches have the identical phase or have a relative phase of $\pi$. In the later case, the integral over the whole surface and with it the coupling efficiency vanishes. In conclusion, we obtain the identical results as discussed in the main text by applying the symmetry argumentation. \newline
Additional information such as the coupling efficiency can be extracted as well, by solving the integrals. However, we do not compare these computed values with the experimental data, since other parameters, such as the antenna coupling efficiency, propagation losses, or the phase matching influence the measured intensities as well, but are unknown and not implemented in this model.

\section{Model for Fig.~4}
\label{sec:ModelFIg4}
Every field distribution at the fundamental frequency in the waveguide can be written as superposition of the  symmetric ($\mathbf{E}_s$) and anti-symmetric ($\mathbf{E}_{as}$) mode:
\begin{equation}
\mathbf{E}_{tot} = a \, \mathbf{E}_s \, + \, b \,  \mathbf{E}_{as} \quad ,
\end{equation}
Rotating the linear polarization of the incoming fundamental laser beam results in
\begin{equation}
 a = \cos \theta \quad \text{and}  \quad b = r \exp(i \phi) \ \sin \theta
\end{equation}
$r$ and $\phi$ take into account that the  excitation and coupling efficiencies of the antenna depend on polarization.
We assume that only the normal component of the field plays an important role \cite{Makitalo2011}, i.e,
\begin{equation}
P^{(2)}_n = \epsilon_0 \,  \chi^{(2)}_{nnn} \, E_n^2   \quad .
\label{eq:p2_nnn}
\end{equation}
Putting everything together, we get
\begin{equation}
 I_{s / as} = scaling_{s / as} \,  \cdot \, \left| \int_{\partial A}  \epsilon_0 \,  \chi^{(2)}_{nnn}  \, \cdot \,  
\left( a \, E_s \, + \, b \, \, E_{as} \right)^2
  \cdot
  \, E_{s / as}(2 \omega)^*
  \, ds \right |^2
\end{equation}
For each emitting mode $E_{s / as}(2 \omega)$ the full integral can be separated in 3 terms which are product of 3 field distributions each. For half of them the integral vanishes due to symmetry. The other three  integrals depend in their value on the spatial shape of the modes and give the complex values $A,B,C$.
\begin{center}
\begin{tabular}{llll} 
\hline
prefactor & fundamental \qquad & second-harmonic \qquad & result \\
\hline
$ \cos^2 \theta$  & $  E_s \, E_s$ & $E_s$  &  $A \,   \cos^2 \theta$    \\
$ r^2  \exp(i 2 \phi) \, \sin^2 \theta$  & $  E_{as} \, E_{as}$ & $E_s$  &    $B \,  r^2 \, \exp(i 2 \phi ) \, \sin^2 \theta  $ \\
$2 r \, \exp(i  \phi)  \, \cos \theta  \, \sin \theta $  & $  E_s \, E_{as}$ & $E_s$  &  0    \\
\hline
$ \cos^2 \theta$  & $  E_s \, E_s$ & $ E_{as}$  &  0    \\
$ r^2  \exp(i 2 \phi) \, \sin^2 \theta$  & $  E_{as} \, E_{as}$ & $ E_{as}$ &   0  \\
$2 r \, \exp(i  \phi)  \, \cos \theta  \, \sin \theta $    & $  E_s \, E_{as}$ & $ E_{as}$ &   $C \, 2 r \, \exp(i  \phi)  \, \cos \theta  \, \sin \theta $  \\
\hline  
\end{tabular}
\end{center}
The values $A$ and $B$ of the integrals can be adsorbed into the fitting parameters $r$ and $\phi$. Moreover, as we need scaling parameters for the absolute values anyway, we can also absorb $C$ in the fitting parameters  $r$ and $\phi$.  Altogether we get
\begin{eqnarray}
I_s & =  & scaling_s    \,  \cdot \, \left(  \cos^4 \theta  \, + \, r^4 \, \sin^4 \theta   \, + \, 2  r^2 \, \cos( 2 \phi) \, \cos^2 \theta \, \sin^2 \theta  \right)  \\
I_{as} & =  &  scaling_{ as}   \,  \cdot \,  4  r^2 \, \cos^2 \theta \,  \sin^2 \theta \, + \, offset \quad .
\label{eq.i_AS}
\end{eqnarray}
In $I_{as}$ an additional $offset$ was added. Here, the model requires a decrease to zero at $\theta = 0\deg$ and $90\deg$ which cannot be fully achieved in the experiment due to structural imperfections, local SHG and a non-perfect mode detector. 

\section{Polar Plots}

In order to validate our model for the second harmonic emission, resulting from a mixed fundamental mode, we perform excitation angle resolved measurements. This allows us to change the amplitude ratio between the two excited fundamental modes while keeping the relative phase constant.
The polarization of the laser beam is tuned using a half wave plate. For each excitation polarization two images are aquired using a horizontal and a vertical analyzer. The fundamental harmonic images are taken in reflection, while the second harmonic images are recorded in transmission. 
For noise reduction all images are filtered by a 3-by-3 pixel median filter and the background is subtracted. In the second harmonic images the background is defined as the mean value from a dark area, while in the fundamental case a line correction is performed to correct artifacts of the NIR camera. Here, the images are additionally smoothed by a Gaussian filter ($\sigma=1.5\, \text{pixel}$, equal to $360\,\text{nm}$ (FWHM 960~nm)). In order to extract the fundamental and second harmonic emission intensities, we apply circular regions of interest to separate the antenna and each of both mode detector ends. The emitted signal from each region is determined by its maximum value. \newline

\begin{figure}
	\includegraphics[width=\columnwidth]{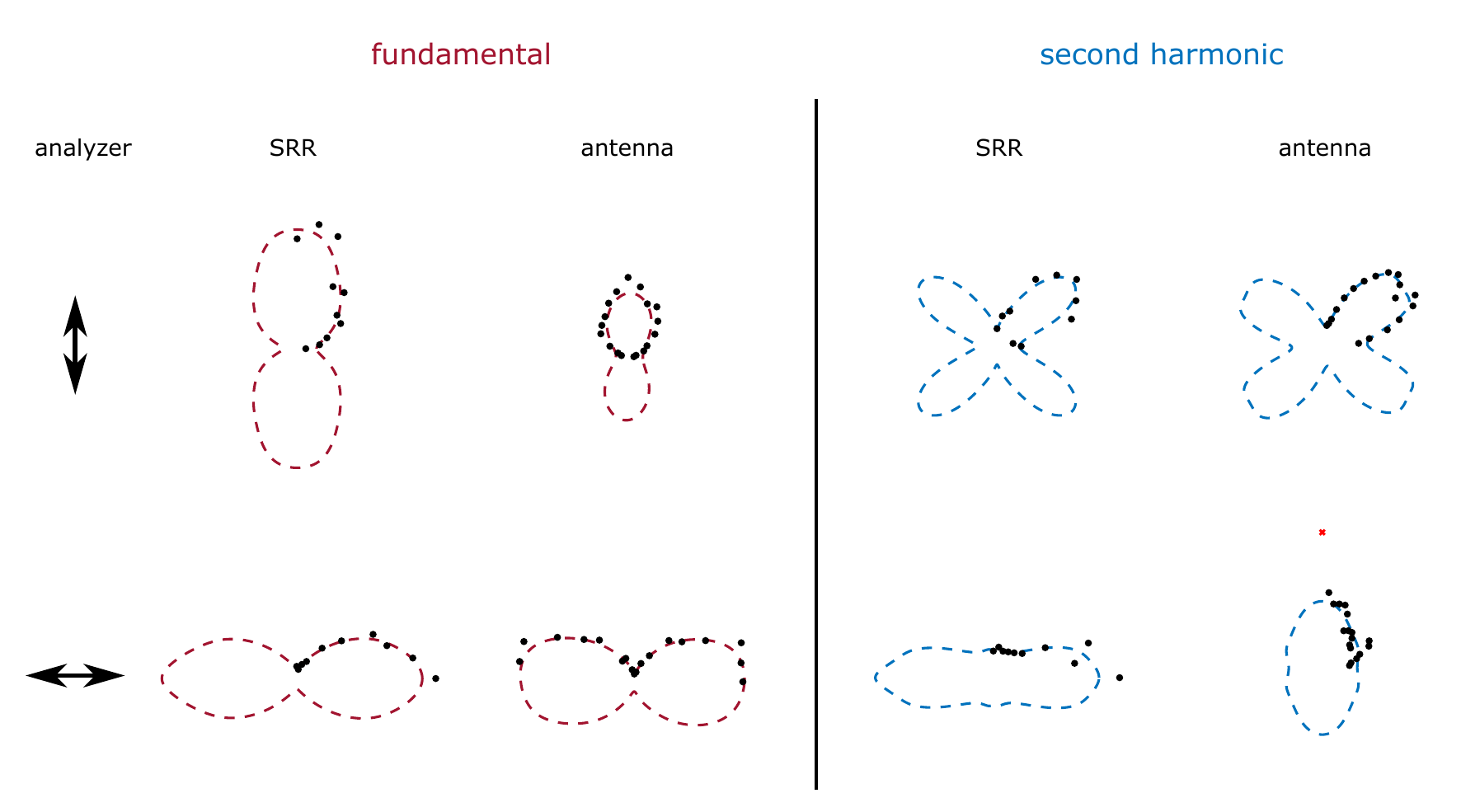}
	\caption{\textbf{Excitation polarization resolved emission from the TWTL's antenna and a split ring resonator.} The fundamental as well as the second harmonic emission is shown for different excitation polarizations. The columns differentiate between wavelength respectively the structure. The rows mark the two analyzer positions. The dashed lines are guides to the eye. The point marked by a red cross represents an excluded data point.
		\label{fig:idea1}}
\end{figure}

We start by discussing the emission properties of the TWTL's antenna and compare them to those of a split ring resonator (SRR), which has a similar geometry. A design comparable in size (arm length 400~nm) ensures that linear and nonlinear scattering properties resemble the ones of the TWTL's antenna without waveguide and mode detector. In analogy to the polar plots presented in the main text (Fig.~4), the excitation polarization dependent emission from the SRR and the TWTL's antenna are shown in Fig.~\ref{fig:idea1}. The columns show the fundamental harmonic and second harmonic emission, the two rows defined the analyzer positions that were used in the detection path.
In the fundamental data we see a strong similarity between the emission pattern from the SRR and the TWTL's antenna for both detection polarizations. Furthermore, the fundamental scattered light follows the polarization properties of the exciting laser beam. 
The second harmonic emission of both structures is also in good agreement. Under the vertical polarization detection, we observe an emission maximum for about $45 \deg$ excitation and minima for $0 \deg$ and $90 \deg$. The deviation in the aspect ratio for the horizontal analyzer position is caused by slightly differing emission properties of the structures,  although the arm length of the SRR is chosen to be off-resonant at the fundamental and the second harmonic frequency.\\ 

\begin{figure}
	\includegraphics[width=\columnwidth]{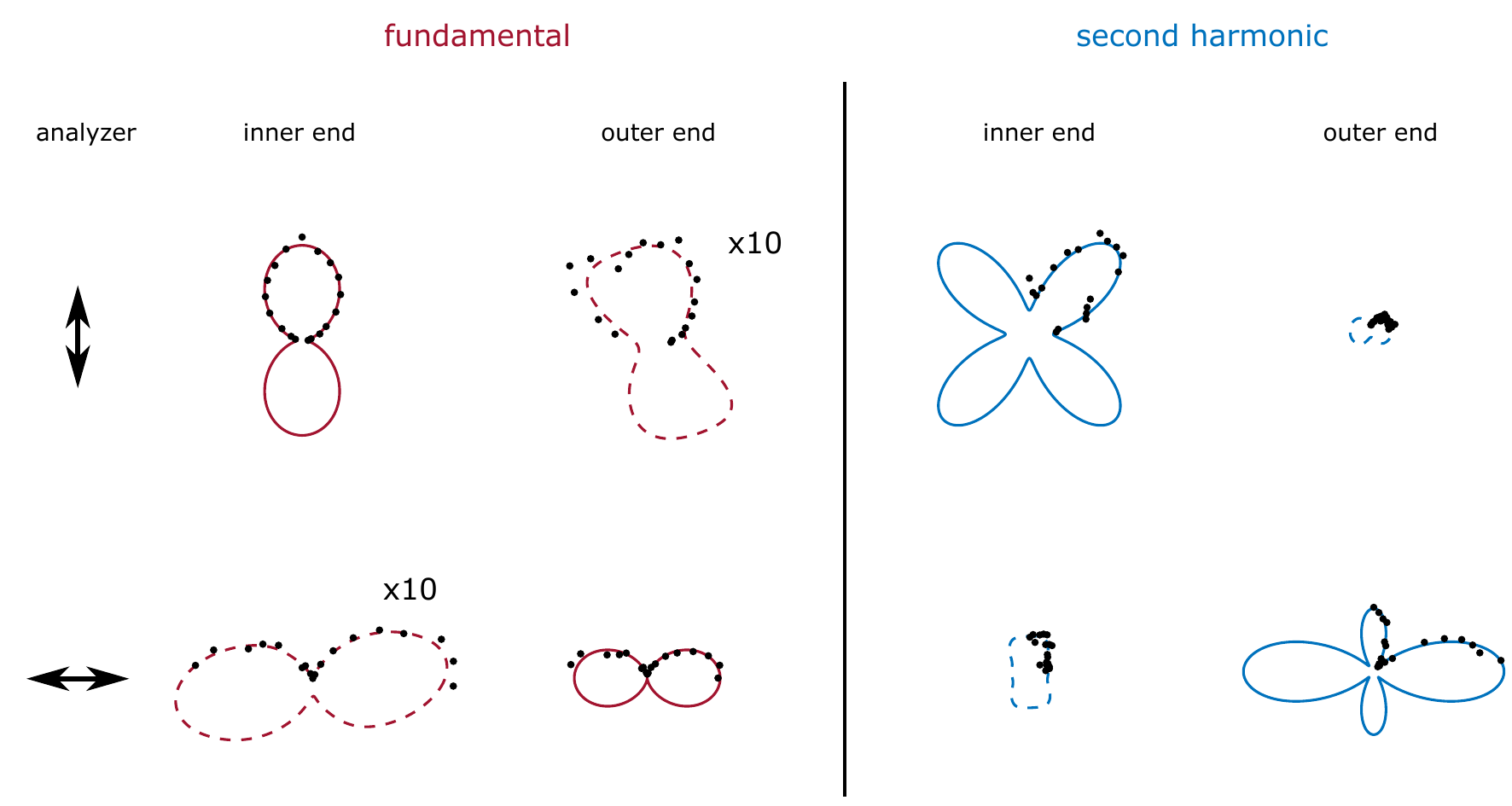}
	\caption{\textbf{Excitation polarization resolved emission from the TWTL's mode detector.} The fundamental as well as the second harmonic emission is shown for different excitation polarizations. The columns differentiate between wavelength respectively the structure. The rows mark the two analyzer positions. The dashed lines are guides to the eye while the solid lines represent fitted physical models. The fundamental data with dashed lines was multiplied by a factor of ten for better visibility. 
		\label{fig:idea2}}
\end{figure}

Finally we investigate the emission properties from the two mode detector ends, shown in Fig.~\ref{fig:idea2}.
The fundamental emission from the mode detector shows the expected behavior as we already verified in the main text for unpolarized detection. As expected, the emission from the inner end of the mode detector is predominantly vertical polarized and the signal from the outer end horizontal. Furthermore, we find at the opposing ends the same emission behavior (shown with dashed lines), but with a factor of $1/10$ less intensity, what can be attributed to the limited mode-splitting quality of the mode detector.

Moving to the second harmonic emission of the mode detector leads us to the right part of Fig.~\ref{fig:idea2}. The diagonal graphs show the SH emission from the anti-symmetric mode (upper left) and symmetric mode (lower right), as already discussed in the main text. Both pattern can be nicely described by our model, taking the SH mode-coupling during propagation into account. However, in contrast to the fundamental harmonic, the second harmonic emission properties of the opposing ends of the mode detector have no similarity with each other. In consequence, the signals shown in the off-diagonal graphs (shown with dashed lines) can not be attributed to the mode splitting properties of the mode detector. Our explanation is locally generated second harmonic light from the fundamental wave, scattered and depolarized at the mode detector ends. Here, the fundamental harmonic field locally generates a nonlinear polarization at the inner and outer end of the mode detector. Due to the curvature and imperfections of the mode-detector geometry, the emitted second harmonic light is strongly depolarized and mainly depends on the incoming fundamental field intensity. In consequence, we find a very good agreement between the overall fundamental harmonic distributions (solid red lines) and the second harmonic background signals (blue dashed lines). It has to be mentioned, that these background signals also underlay the fitted second harmonic data (solid blue lines). In case of the SH-antisymmetric mode (upper left) this background signal is part of the offset, implemented in the model for $ I_{as} $ (see equation~\ref{eq.i_AS}). For the SH-symmetric mode (lower right), an offset is not necessary or is fitted to almost zero, respectively. Based on the good agreement between model and experiment as well as the small background signals we can conclude, that the locally generated second harmonic plays an minor role and can be almost neglected in the overall measured signals.


%